\documentclass[11pt,a4paper]{scrartcl}

\usepackage[centertags]{amsmath} 
\usepackage{color} 
\usepackage{amsfonts}
\usepackage{amssymb}
\usepackage{amsthm}
\usepackage{caption}
\usepackage{booktabs}
\usepackage{enumitem}
\usepackage{threeparttable}
\usepackage{graphicx}   
\usepackage[numbers]{natbib}
\usepackage[left=3cm,right=3cm,top=3.5cm,bottom=4cm]{geometry}

\newcommand{\ab}{\boldsymbol{a}}
\newcommand{\xb}{\boldsymbol{x}}
\newcommand{\ub}{\boldsymbol{u}}

\newcommand{\kb}{\boldsymbol{k}}
\newcommand{\lb}{\boldsymbol{l}}

\newcommand{\wb}{\boldsymbol{w}}

\newcommand{\mub}{\boldsymbol{\mu}}

\newcommand{\lambdab}{\boldsymbol{\lambda}}

\newcommand{\Ab}{\boldsymbol{A}}
\newcommand{\Bb}{\boldsymbol{B}}

\newcommand{\Qb}{\boldsymbol{Q}}
\newcommand{\Rb}{\boldsymbol{R}}
\newcommand{\Pb}{\boldsymbol{P}}

\newcommand{\Vb}{\boldsymbol{V}}

\newcommand{\Ib}{\boldsymbol{I}}

\newcommand{\N}{\mathbb{N}}

\newcommand{\R}{\mathbb{R}}

\newcommand{\Ac}{\mathcal{A}}

\newcommand{\Kc}{\mathcal{K}}

\newcommand{\Xc}{\mathcal{X}}

\newcommand{\Uc}{\mathcal{U}}

\newcommand{\Tc}{\mathcal{T}}
\newcommand{\Pc}{\mathcal{P}}
\newcommand{\Fc}{\mathcal{F}}

\newcommand{\Ic}{\mathcal{I}}

\newcommand{\Vc}{\mathcal{V}}

\newcommand{\interior}{\mathrm{int}}

\newtheorem{thm}{Theorem}

\newtheorem{assum}{Assumption}
\newtheorem{prop}[thm]{Proposition}

\begin{document}

\begin{center}
% Title
{\LARGE
\textbf{\textsf{Novel convex decomposition \\[2mm]
of piecewise affine functions}}
}

\renewcommand{\thefootnote}{$\dagger$} 

\vspace{5mm}
{
Nils Schl\"uter\footnotemark[1] and Moritz Schulze Darup\footnotemark[1]
}
\vspace{2mm}

  \footnotetext[1]{Nils Schl\"uter and Moritz Schulze Darup are with the Control and Cyberphysical Systems Group, Department of Mechanical Engineering, TU Dortmund University, Germany. E-mails: \texttt{\{nils.schlueter,moritz.schulzedarup\}@tu-dortmund.de}
  }
\end{center}

\vspace{1mm}
\paragraph{Abstract.}
In this paper, we present a novel approach to decompose a given piecewise affine (PWA) function into two convex PWA functions. Convex decompositions are useful to speed up or distribute evaluations of PWA functions. Different approaches to construct a convex decomposition have already been published. However, either the two resulting convex functions have very high or very different complexities, which is often undesirable, or the decomposition procedure is inapplicable even for simple cases. Our novel methodology significantly reduces these drawbacks in order to extend the applicability of convex decompositions.


\paragraph{Preamble.}
This paper is a reprint of a contribution (``late breaking result'') to the 21st IFAC World Congress 2020.

%===============================================================================
\vspace{5mm}
\section{Motivation and overview}
PWA functions arise frequently in automatic control and elsewhere. A popular example is explicit model predictive control~\cite{Bemporad2002}. Classically, the evaluation of a PWA function $f(\xb)$ for a given $\xb$ in its domain is two-stage. First, the segment of $f$ that belongs to $\xb$ is identified. Second, the corresponding affine function is evaluated. More efficient or distributed evaluations of PWA functions can be realized by rewriting $f$ as the difference of two convex PWA functions. In fact, convexity of PWA functions can be exploited to reduce memory consumption and computational effort significantly~\cite{nguyen2017convex}. Another application of convex decompositions is DC programming~\cite{horst1999dc} that allows to globally solve certain non-convex optimization problems.

While convex decompositions are useful, their construction is typically cumbersome. For instance, the approach presented in \cite{Kripfganz1987} decomposes $f$ into two convex PWA functions $g$ and $h$, where especially the construction of $h$ is numerically  demanding. A simpler construction is proposed in \cite{Hempel2015}, but the procedure is often not applicable. In this paper, we present a novel convex decomposition that reduces the weaknesses of both existing approaches while maintaining their strengths. To this end, we summarize the existing approaches in Section \ref{sec:Existing}. Our novel method is presented in Section~\ref{sec:novel} and illustrated with an example in Section~\ref{sec:casestudy}. Finally, conclusions are given in Section~\ref{sec:conclusions}.

\section{Existing convex decompositions}
\label{sec:Existing}
\begin{figure}[htp]
	\centering
	\includegraphics[trim=4.5cm 10cm 4.5cm 10.2cm, clip=true,width=0.7\textwidth]{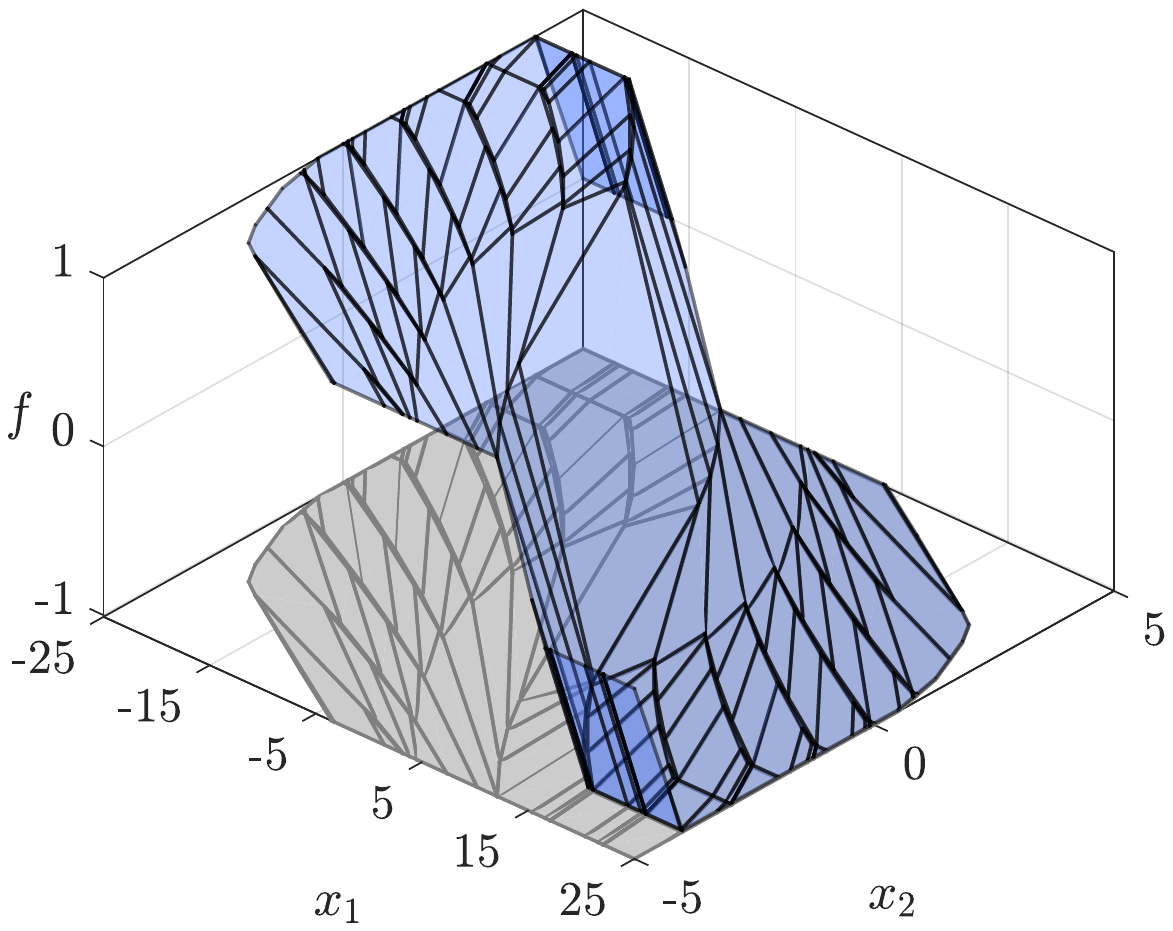}
		\caption{Illustration of the PWA function $f$ resulting from the MPC example in Section~\ref{sec:casestudy} and $N=10$.}
	\label{fig:PWA}
\end{figure}

Throughout the paper, we focus on the decomposition of a given continuous PWA function $f:\R^n \rightarrow \R$ of the form
\begin{equation}
\label{eq:f}
f(\xb):=\left\{ \begin{array}{cl}
\ab_1^\top \xb + b_1 & \quad  \text{if}\,\, \xb \in \Xc_1, \\
\vdots \\
\ab_s^\top \xb + b_s & \quad  \text{if}\,\, \xb \in \Xc_s, \\
\end{array}\right.
\end{equation}
into two convex PWA  functions $g$ and $h$ such that
\begin{equation}
\label{eq:f=g-h}
f(\xb)= g(\xb)-h(\xb)
\end{equation}
holds for every $\xb$ in the domain $\Fc := \bigcup_{i=1}^s \Xc_i$
of $f$. In this context, the partition $\{ \Xc_i \}_{i=1}^s$ (often abbreviated as $\{\Xc_i\}$) is assumed to satisfy the following conditions.
\begin{assum}
\label{assum:Xij}
The sets $\Xc_i$ are polyhedral, convex and offer $\interior(\Xc_i)\neq \emptyset$ (nonempty interiors) as well as $\interior(\Xc_i) \cap \interior(\Xc_j) = \emptyset$ for every $i\neq j$ (pairwise disjoint interiors).
\end{assum}
We note, however, that sets $\Xc_i$ and $\Xc_j$ may have overlapping boundaries. In such cases, continuity of $f$ requires $\ab_{i}^{\top} \xb + b_i = \ab_{j}^\top \xb + b_j$ whenever $\xb \in \Xc_i \cap \Xc_j$. For completeness, we finally note that $\ab_i \in \R^n$, $b_i \in \R$, and $s \in \N$ with $s$ referring to the number of segments in~\eqref{eq:f}. An example of a function $f$ as defined above is shown in Figure~\ref{fig:PWA}. It is well known that a decomposition of the form~\eqref{eq:f=g-h} is not unique but in principle always possible \citep{Kripfganz1987}. Two existing approaches will be discussed next.

\subsection{Decomposition via convex folds}
The first approach builds on the constructive decomposition proof in \cite{Kripfganz1987}. The underlying idea is to collect all convex folds of $f$ and to use them in a certain way to construct $g$.  More formally, let 
$$
\Ic := \big\{ (i,j)\in \{1,\dots,s\}^2 \,\big|\, \dim( \Xc_i \cap \Xc_j) = n-1,\,  i<j
\big\} 
$$
collect index pairs of neighboring polyhedra $\Xc_i$ and $\Xc_j$ that share a common facet. Further, let 
\begin{equation}
\label{eq:V}
\Vc:= \big\{ (i,j) \in \Ic \,\big|\,  \ab_i^\top \xb + b_i > \ab_j^\top \xb + b_j, \, \xb \in \Xc_i \setminus \Xc_j \big\} \text{,}
\end{equation}
denote the subset of $\Ic$ that collects facets on which $f$ features a convex fold. Then,
\begin{equation}
\label{eq:collectconvex}
g(\xb):= \sum_{(i,j) \in \Vc} \max \{ \ab_i^\top \xb + b_i ,\ab_j^\top \xb + b_j \} 
\end{equation}
is obviously a convex function since the maximum of affine functions is convex and since sums preserve convexity. More interestingly, the function 
\begin{equation}
\label{eq:h=g-f}
h(\xb):= g(\xb)-f(\xb)
\end{equation}
is convex \cite[Lem.~1]{Kripfganz1987}.  Since \eqref{eq:f=g-h} holds by construction, $g$ and $h$ indeed form a convex decomposition of $f$. 

\begin{figure}[tp]
	\centering
	\includegraphics[trim=3.5cm 10cm 4.5cm 10.2cm, clip=true,width=0.75\linewidth]{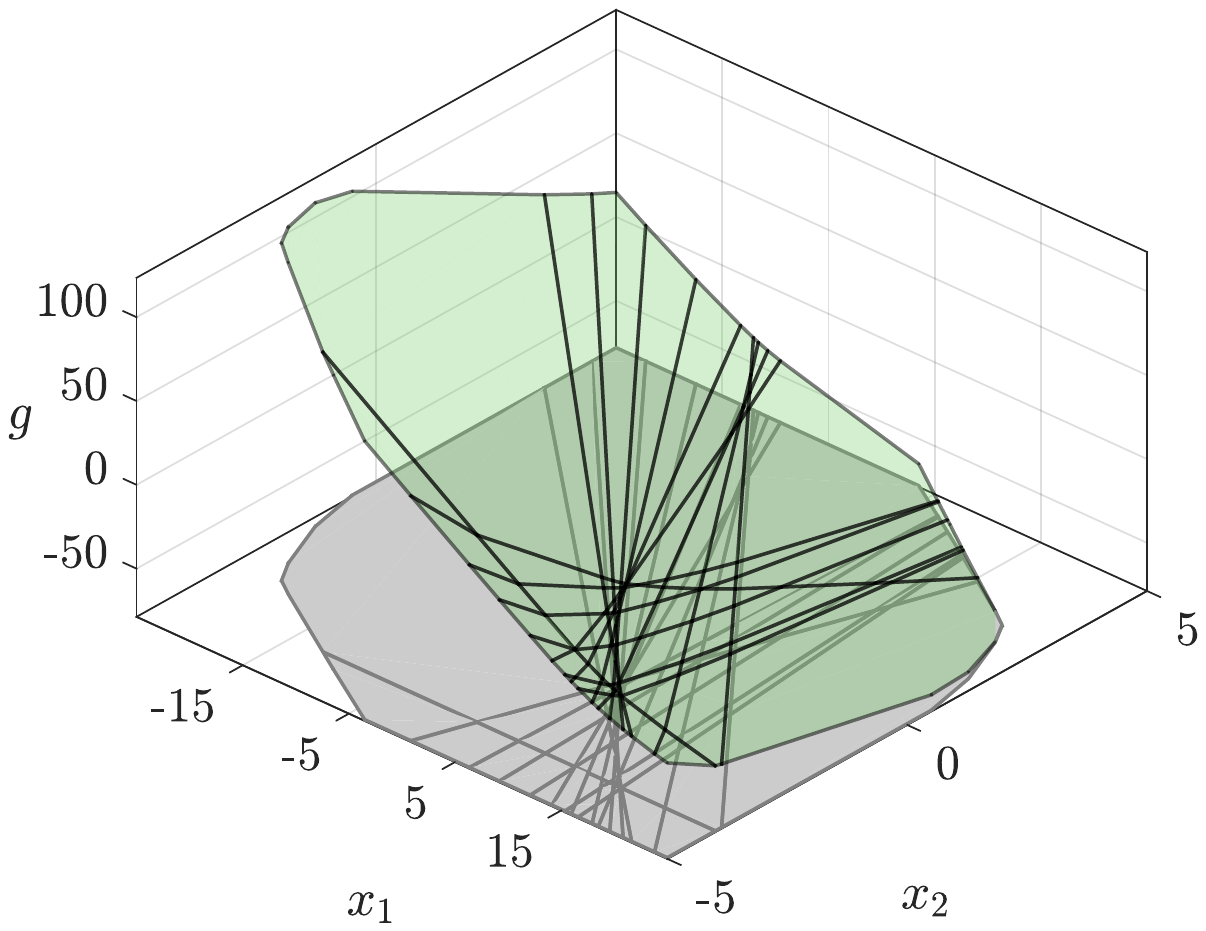} 
	\caption{Illustration of \eqref{eq:collectconvex} evaluated for the PWA control law shown in Figure~\ref{fig:PWA}.}
	\label{fig:AnitaG}
\end{figure}

\begin{figure}[tp]
	\centering
	\includegraphics[trim=3.5cm 10cm 4.5cm 10.2cm, clip=true,width=0.75\linewidth]{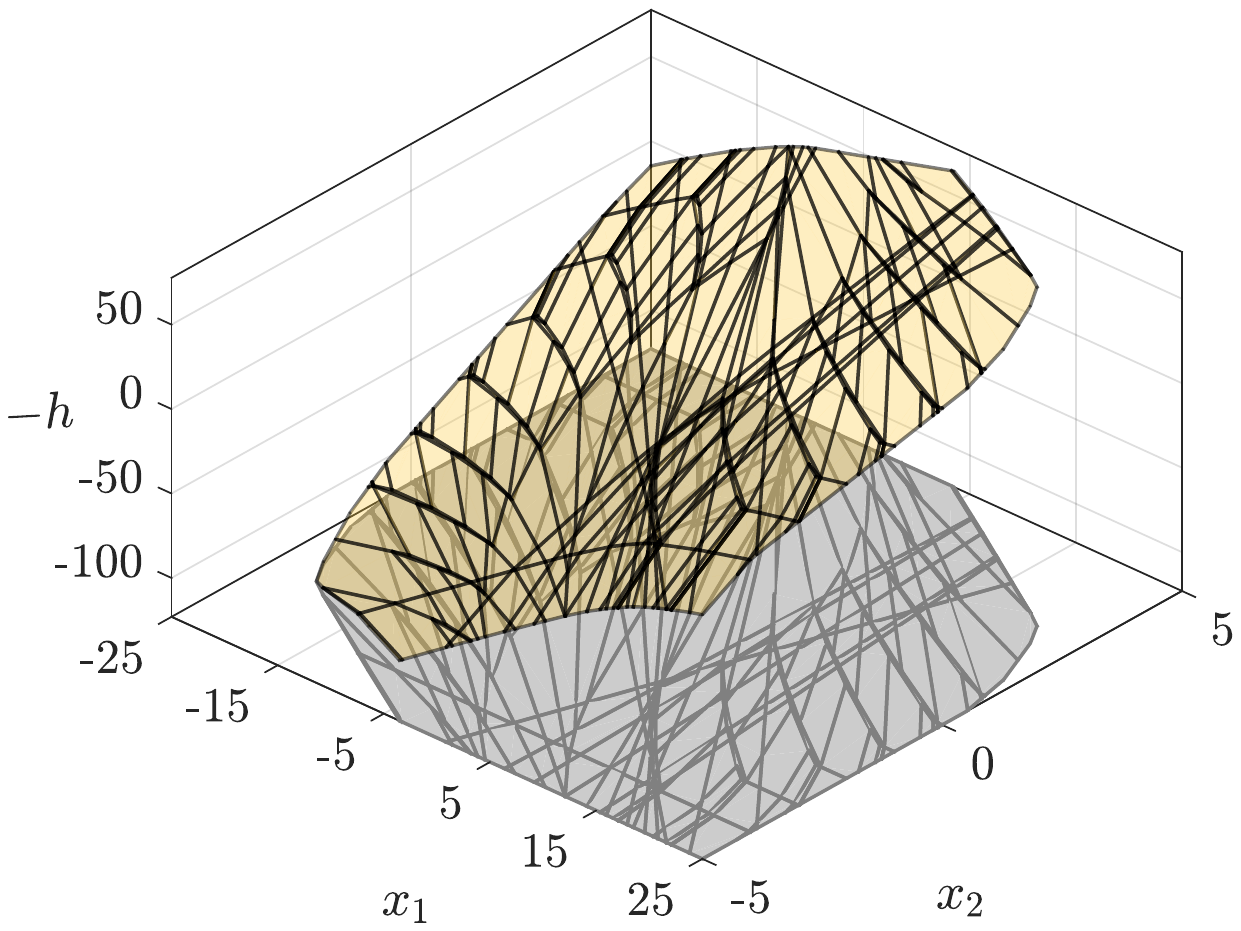}
	\caption{Illustration of \eqref{eq:h=g-f} evaluated for the PWA control law shown in Figure~\ref{fig:PWA}.}
	\label{fig:AnitaH}
\end{figure}

While the decomposition is elegant from a mathematical point of view, it is (computationally) demanding to express $g$ and $h$ in a form similar to~\eqref{eq:f}. Regarding $g$, we note that every summand $\max \{ \ab_i^\top \xb + b_i ,\ab_j^\top \xb + b_j \} $ refers to a convex PWA function with two segments implicitly defined on the two halfspaces 
\begin{equation}
\label{eq:halfspaces}
\ab_i^\top \xb + b_i \geq \ab_j^\top \xb + b_j \quad \text{and} \quad
\ab_i^\top \xb + b_i \leq \ab_j^\top \xb + b_j,
\end{equation}
respectively. The superposition (or summation) of all these one-folded functions leads to a convex PWA function as in Figure~\ref{fig:AnitaG}. Note that the underlying partition results from ``cutting'' $\Fc$ using every separating hyperplane $\ab_i^\top \xb + b_i = \ab_j^\top \xb + b_j$ induced by~\eqref{eq:halfspaces} for $(i,j)\in\Vc$. Overlaying the resulting partition for $g$ with the original partition $\{\Xc_i\}$ of $f$ leads to another partition that allows to express $h$ as a PWA function. In fact, since $g$ and $f$ are affine on every subset of the latter partition, also $h$ is affine there as a consequence of \eqref{eq:h=g-f}. Unfortunately, the partition of $h$ is often significantly finer (i.e., it consists of more polyhedra) than the ones of $f$ and $g$. This effect is, for example, apparent from Figure~\ref{fig:AnitaH}.

\subsection{Optimization-based decomposition}
\label{subsec:HempelsApproach}
As proposed in \cite{Hempel2015}, a convex decomposition can also be constructed optimization-based. In contrast to the previous approach, the optimization-based decomposition yields functions $g$ and $h$, which are defined on the same partition $\{\Xc_i\}$ as $f$. In other words, the functions $f$, $g$, and $h$ will all be affine on each polyhedron $\Xc_i$. The corresponding affine segments of $g$ and $h$ will be denoted with $\kb_{i}^{\top} \xb + c_i$ and $\lb_{i}^{\top} \xb + d_i$, respectively. A decomposition satisfying~\eqref{eq:f=g-h} then requires
\begin{equation}
\label{eq:decomp}
\ab_{i}=\kb_{i}-\lb_{i} \quad  \text{and} \quad
b_{i}=c_{i}-d_{i} 
\end{equation}
for every $i\in \{1,\dots,s\}$. It remains to enforce convexity of $g$ and $h$. To this end, for every $(i,j)\in \Ic$, we consider the inequality constraints
\begin{subequations}
\label{eq:decompconstr}
\begin{equation}
\label{eq:decompconstr1}
\kb_{i}^{\top} \xb+c_{i}\geq \kb_{j}^{\top } \xb+c_{j} \quad \text{and} \quad
\lb_{i}^{\top} \xb+d_{i}\geq \lb_{j}^{\top } \xb+d_{j}
\end{equation}
for every $\xb \in \Xc_i$ as well as
\begin{equation}
\label{eq:decompconstr2}
\kb_{i}^{\top} \xb+c_{i}\leq \kb_{j}^{\top } \xb+c_{j} \quad \text{and} \quad
\lb_{i}^{\top} \xb+d_{i}\leq \lb_{j}^{\top } \xb+d_{j}
\end{equation}
\end{subequations}
for  every $\xb \in \Xc_j$. Obviously, the combination of the first condition in~\eqref{eq:decompconstr1} and \eqref{eq:decompconstr2} implies $\kb_{i}^{\top} \xb+c_{i} = \kb_{j}^{\top } \xb+c_{j}$ for every $\xb \in \Xc_i \cap \Xc_j$, i.e., continuity of $g$. Analogously, continuity of $h$ is ensured. We further note that, in contrast to~\eqref{eq:V}, strict convexity is not required in \eqref{eq:decompconstr}. 

\begin{figure}[hp]
	\centering
	\includegraphics[trim=4.2cm 11cm 4.7cm 11cm, clip=true,width=0.72\linewidth]{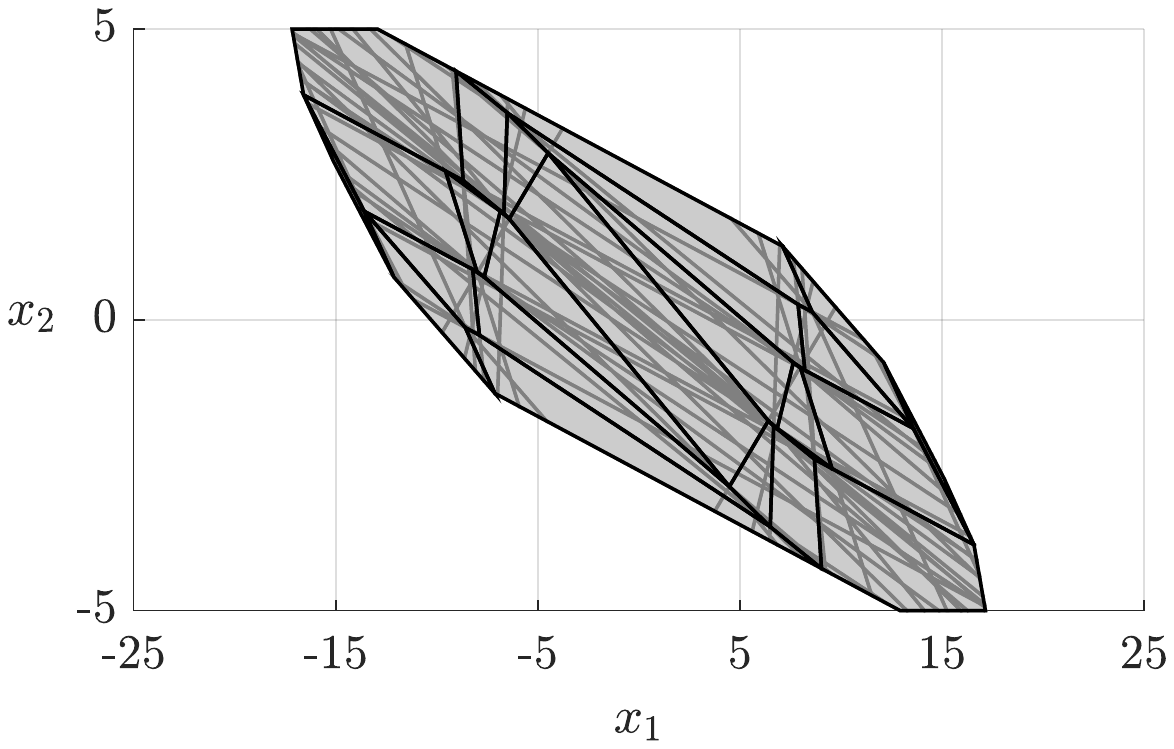}
	\caption{Hyperplane arrangement (dark gray partition) applied to a non-regular partition (black) resulting from the MPC example in Section \ref{sec:casestudy} and $N=3$.}
	\label{fig:hyperplane}
\end{figure}

Assuming half-space representations of the subsets $\Xc_i$ are at hand, i.e., $\Xc_i=\{\xb \in \R^{n} \,|\, \Vb_i \xb \leq \wb_i\}$, \eqref{eq:decompconstr} can be efficiently verified using Farkas's lemma. Conditions \eqref{eq:decompconstr} are satisfied if and only if there exist (Lagrange multipliers) $\lambdab_{ij}$, $\mub_{ij}$, $\lambdab_{ji}$, and $\mub_{ji}$ of appropriate dimensions such that
\begin{subequations}
\label{eq:convconstr}
\begin{align}
\boldsymbol{0} &\leq \lambdab_{i j},\,\, 
\Vb_{i}^{\top} \lambdab_{ij}=\left(\kb_{j}-\kb_{i}\right)^{\top}\!\!,\,\,
\wb_{i}^{\top}\lambdab_{ij} \leq c_{i}-c_{j}\\
\boldsymbol{0} &\leq \mub_{i j},\,\, 
\Vb_{i}^{\top}\mub_{ij}=\left(\lb_{j}-\lb_{i}\right)^{\top}\!\!,\,\,\,\,
\wb_{i}^{\top}\mub_{ij} \leq d_{i}-d_{j} \text{,}\\
\boldsymbol{0} &\leq \lambdab_{ji},\,\, 
\Vb_{j}^{\top} \lambdab_{ji}=\left(\kb_{i}-\kb_{j}\right)^{\top}\!\!,\,\,
\wb_{j}^{\top}\lambdab_{ji} \leq c_{j}-c_{i},\\
\boldsymbol{0} &\leq \mub_{ji},\,\, 
\Vb_{j}^{\top}\mub_{ji}=\left(\lb_{i}-\lb_{j}\right)^{\top}\!\!,\,\,\,\,
\wb_{j}^{\top}\mub_{ji} \leq d_{j}-d_{i}.
\end{align}
\end{subequations}
Now, any feasible solution to \eqref{eq:decomp} and \eqref{eq:convconstr} provides a valid decomposition of $f$ into two convex PWA functions. The feasibility problem can be extended by a user-defined cost function or additional constraints in order to promote certain features of $g$ and~$h$. For example, minimizing the quadratic cost function
$$
\sum_{i=1}^{s}\kb_{i}^{\top} \kb_{i}+c_{i}^2+ \lb_{i}^{\top} \lb_{i} +  d_{i}^2
$$
subject to \eqref{eq:decomp} and \eqref{eq:convconstr} promotes small coefficients (absolute values) for $g$ and $h$.

Unfortunately, a severe drawback of this decomposition is that feasibility of the optimization problem requires regularity of the partition $\{\Xc_i\}$ (see \cite[page 53]{de2010triangulations} for details), which is often not fulfilled even for simple partitions. To regularize a non-regular partition, hyperplane arrangement as proposed in \cite{nguyen2017convex} can be used. Here, the hyperplanes defining each polyhedron $\Xc_i$ are extended to the boundary of $\Fc$. If polyhedrons intersect these extended hyperplanes, they are split. The result is a highly refined partition as illustrated in Figure~\ref{fig:hyperplane} for an example. Due to the high number of polyhedrons, illustrating this method for finer partitions ($N\geq 5$, see Section~\ref{sec:casestudy}) is meaningless.

\section{Novel convex decomposition}
\label{sec:novel}
As an intermediate summary, the approach in \cite{Kripfganz1987} typically provides a simple partition (and construction) for $g$ and a complex one for $h$. The approach in \cite{Hempel2015} allows for user-defined designs of $g$ and $h$ but the underlying optimization problem is often not feasible without additional regularization strategies. In the following, we present a novel optimization-based decomposition scheme that is always applicable and that provides functions $g$ and $h$ with identical complexities.

As a preparation, we introduce the set 
$$
\Ac:= \{ (i,j) \in \Ic \,|\,  \ab_i^\top \xb + b_i < \ab_j^\top \xb + b_j, \, \xb \in \Xc_i \setminus \Xc_j \}
$$
that, analogously to~\eqref{eq:V}, collects all concave folds of $f$.
Based on this set, one is tempted to construct $h$ as
\begin{equation}
\label{eq:hWrongGuess}
h(\xb):=-\sum_{(i,j) \in \Ac} \min \{\ab_i^\top \xb+ b_i ,\ab_j^\top \xb + b_j \}
\end{equation}
in analogy to \eqref{eq:collectconvex}. While such an $h$ would indeed be convex, condition \eqref{eq:f=g-h} would not be satisfied in general.
However, it is easy to see that the combined partitions induced by~\eqref{eq:collectconvex} and \eqref{eq:hWrongGuess} are always regular. In fact, both can be considered as a hyperplane arrangement for the convex respectively concave folds of $f$. Our simple idea for a novel decomposition is to consider this combined partition for an optimization-based decomposition. 
More precisely, let
$$
(i_1,j_1),\dots,(i_p,j_p) 
$$ 
denote the $p:=|\Vc|+|\Ac|$ index pairs in $\Vc \cup \Ac$.
Now, for any $k \in \{1,\dots,2^p\}$, let $\beta_1, \dots, \beta_p \in \{0,1\}$ express the unique binary representation satisfying
$$
k= 1+\sum_{d=1}^{p} \beta_d 2^{d-1}.
$$
Then, we define the $k$-th subset of the novel partition as
\begin{align*}
\Pc_k := \Big\{ \xb \in \Fc \,|\,  & (-1)^{\beta_1} \left((\ab_{i_1} -\ab_{j_1})^\top \xb +b_{i_1}- b_{j_1}\right) \geq 0,\\
& \qquad \qquad \quad \vdots \\
& (-1)^{\beta_p} \left((\ab_{i_p} -\ab_{j_p})^\top \xb +b_{i_p}- b_{j_p}\right) \geq 0 \Big\}.
\end{align*}
Typically, many of these sets are empty or of lower dimension than $n$. Hence, we consider only those subsets with non-empty interiors, i.e., the sets $\Pc_k$ with 
$$
k\in \Kc := \{ k \in \{1,\dots,2^p\} \,|\, \interior(\Pc_k) \neq \emptyset \}.
$$
The sets $\Pc_k$ reflect all combinations of the halfspaces~\eqref{eq:halfspaces}
for all $(i,j) \in \Vc \cup \Ac$ intersected with the set $\Fc$.
Hence, the following proposition holds by construction.
\setcounter{thm}{0}
\begin{prop}
\label{prop:PartitionEqualsX}
Let $\Fc$, $\Pc_k$, and $\Kc$ be as above. Then, $\{\Pc_k\}$ is a regular partition and
$\Fc= \bigcup_{k \in \Kc} \Pc_k$.
\end{prop}

We note, at this point, that $\Kc$ can be efficiently computed without an extensive search over all $2^p$ combinations, e.g., by using binary search trees. Next, before presenting our optimization-based decomposition, we define the function $f^\prime:=\Fc \rightarrow \R$ segment-wise, for every $k \in \Kc$, as
$$
f^\prime(\xb):= \ab_{l_k}^{\top} \xb + b_{l_k} \quad \text{whenever} \quad \xb \in \Pc_k,
$$
where $l_k$ is an arbitrary but fixed ${l_k} \in \{1,\dots,s\}$ satisfying $\interior(\Xc_{l_k}) \cap \interior(\Pc_k) \neq \emptyset$. Such an $l_k$ exists for every $k \in \Kc$ as a result of Assumption~\ref{assum:Xij},  $\interior(\Pc_k) \neq \emptyset $, and Proposition~\ref{prop:PartitionEqualsX}.
Not surprisingly, $f^\prime$ is equivalent to $f$ as specified in the following proposition.

\begin{prop}
\label{prop:ffEqual}
Let $f$ and $f^\prime$ be defined as above. Then,
$$
f(\xb)=f^\prime(\xb)
$$
for every $\xb \in \Fc$.
\end{prop}

We omit a formal proof of Proposition~\ref{prop:ffEqual} due to space restrictions and concentrate on the application of the results above. In this context, we simply apply the optimization-based decomposition from Section~\ref{subsec:HempelsApproach} to the function $f^\prime$ defined on $\{\Pc_k\}$. Since $\{\Pc_k\}$ is regular by construction, the corresponding optimization problem is always feasible and since $f^\prime$ is equivalent to $f$, we obtain a valid decomposition for $f$ with identical complexities of $g$ and $h$.

\section{Case study for explicit MPC}
\label{sec:casestudy}
We study an explicit model predictive controller (MPC) to illustrate our novel decomposition and to compare it with the existing ones. In this context we recall that explicit MPC for linear systems with polyhedral constraints and quadratic performance criteria is known to result in PWA control laws~\citep{Bemporad2002}.

For simplicity, the double integrator dynamics
\begin{equation*}
\label{eq:linearSystem}
\xb(k+1)=\Ab \xb(k) + \Bb u(k)=
\begin{pmatrix}
1 & 1 \\ 0 & 1
\end{pmatrix} \xb(k) + \begin{pmatrix}
0.5 \\1
\end{pmatrix} \ub(k)
\end{equation*} 
are considered with the state and input constraints
\begin{align*}
\xb(k) \in \Xc&:=\{ \xb \in \R^2 \,|\, |x_1| \leq 25, \, |x_2| \leq 5 \}\quad \text{and} \\  
u(k) \in \, \Uc&:=\{ u \in \R \, |\, |u| \leq 1 \}\text{.} 
\end{align*}
MPC then builds on solving the optimal control problem 
$$
V(\xb) := \!\!\min_{\substack{\tilde{\xb}(0),...,\tilde{\xb}(N)\\ \tilde{\ub}(0),...,\tilde{\ub}(N-1)}} \!\!  \|\tilde{\xb}(N)\|_{\Pb}^2  + \!\!\sum_{\kappa=0}^{N-1} \|\tilde{\xb}(\kappa)\|_{\Qb}^2 + \|\tilde{\ub}(\kappa)\|_{\Rb}^2
$$
\begin{align}
\label{eq:OCP}
\text{s.t.} \qquad\quad \tilde{\xb}(0) & = \xb, &&  \\
\nonumber
\tilde{\xb}(\kappa+1)&=\Ab\,\tilde{\xb}(\kappa) + \Bb \tilde{\ub}(\kappa), && \forall \kappa \in \{0,\dots,N-1\}  \\
\nonumber
\tilde{\xb}(\kappa) &  \in \Xc, && \forall \kappa \in \{0,\dots,N-1\} \\
\nonumber
\tilde{\ub}(\kappa) & \in \Uc, && \forall \kappa \in \{0,\dots,N-1\} \\
\nonumber
\tilde{\xb}(N) & \in \Tc &&
\end{align}
in every time step for the current state $\xb = \xb(k)$.
Here, $N$ refers to the prediction horizon, $\Qb$, $\Rb$, and $\Pb$ are weighting matrices, and $\Tc$ is a terminal set. The control action at time $k$ refers to the first element of the optimal control sequence, i.e., $\ub(k)=\tilde{\ub}^{*}(0)$. For our numerical benchmark, we choose $N \in \{1,5,10,15\}$,  $\Qb=\Ib$ and $R=1$. The (positive definite) matrix $\Pb$ is the solution to the  discrete-time algebraic Riccati equation. The set $\Tc$ is chosen as the largest subset of $\Xc$, where the linear quadratic regulator can be applied without violating constraints. It is well known that~\eqref{eq:OCP} can be rewritten as a parametric quadratic program that admits a PWA solution in its parameter~\citep{Bemporad2002}. As a consequence, also the control law $f(\xb):=\tilde{\ub}^{*}(0)$ is PWA. Next, we apply the two existing decompositions and our novel approach to this $f(\xb)$, which is illustrated in Figure~\ref{fig:PWA} for the example at hand and $N=10$.

With regard to practical applications, we are mainly interested in the complexity of the resulting functions $g$ and $h$. We measure their complexity by counting the number of polyhedrons forming the underlying partitions. These numbers are compared with the number of segments $s$ of $f$ for different $N$. Numerical results are given in Table~\ref{table:complexity}. 

\setcounter{table}{0}
\begin{table}[hp]
\centering
	\caption{Number of polyhedrons for resulting partitions with varying $N$}
	\label{table:complexity}
\begin{threeparttable}
\begin{tabular}{l l l l l l}  
	\toprule
	    & $N=1$ &$5$ & $10$ & $15$ \\
	\midrule
\qquad initial partition 				   & $7$      & $75$  		& $223$		& $293$      \qquad\qquad\\
\qquad via convex folds$^\dagger$  & $14,19$  & $103,298$   & $105,581$ & $106,697$  \qquad\qquad\\
\qquad optimization-based$^\star$         & $7$      & $4353$		& $22638$  	& $26786$     \qquad\qquad   \\
\qquad novel decomposition 			   & $33$     & $331$   	& $339$   	& $347$    \qquad\qquad  \\
%  $N=3$ & $29$ & $37,94$  & $685$  & $113$
%  $N=7$ & $147$ & $104,448$ & $-$ & $331$
	\bottomrule
\end{tabular}
\begin{tablenotes}
	\item[$\dagger$] complexity of $g$ and $h$, respectively
	\item[$^{\star}$] for $N>1$ hyperplane arrangement is used for regularization
\end{tablenotes}
\end{threeparttable}
\end{table}

\begin{figure}[h]
	\centering
	\includegraphics[trim=4.5cm 10.3cm 4cm 10.9cm, clip=true,width=0.75\linewidth]{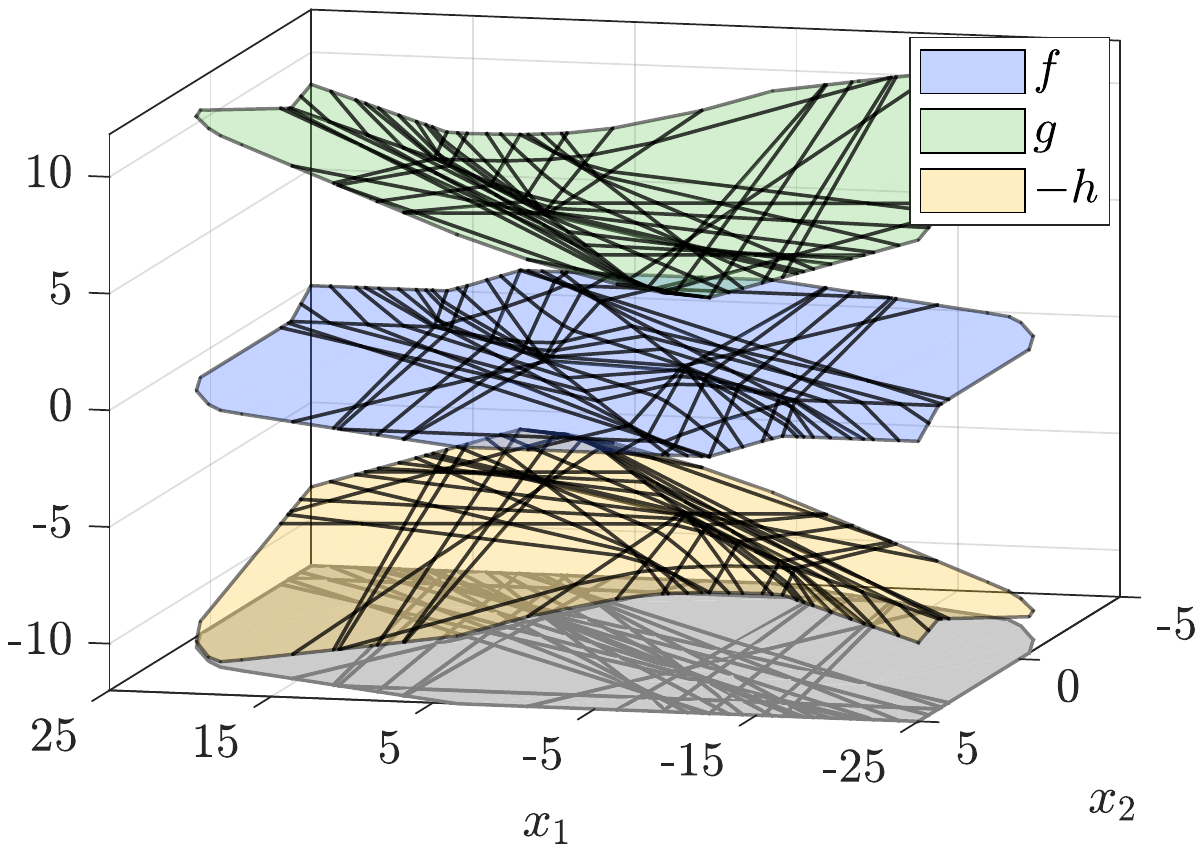}
	\caption{Novel decomposition for $f$ as in Figure~\ref{fig:PWA}.}
	\label{fig:noveldecomp}
\end{figure}

As apparent from the table, we obtain different complexities for $g$ and $h$ using the decomposition from~\cite{Kripfganz1987}. Moreover, the approach from \cite{Hempel2015} is, without hyperplane arrangement, only applicable for the trivial case $N=1$. In all other cases, i.e., for $N>1$, a regularization has to be applied. Following the hyperplane arrangement approach in~\cite[Alg.  4]{nguyen2017convex}, we obtain partitions with the listed complexities. Finally, the complexity of the partition $\{\Pc_k\}$ underlying our novel decomposition is given in the last row of Table~\ref{table:complexity}. An illustration for $N=10$ can be found in Figure~\ref{fig:noveldecomp}. It can be seen that every method refines the initial partition $\left\{\Xc_i\right\}$. A decomposition via convex folds leads to significantly more complex partitions for $h$. Due to hyperplane arrangement the partition related to the optimization-based approach gains rapidly in complexity, rendering the method impractical for complex initial partitions. Our approach provides equal and moderate complexities for both functions $g$ and $h$. Interestingly, for $N=15$, we obtain an accumulated complexity of $2 \times 347=694$ that is even smaller than $106+697 = 803$ as for the approach from~\cite{Kripfganz1987}.

As initially mentioned, convex decompositions can be used to speed up the evaluation of $f$. To see this, note that
\begin{align}
\label{eq:maxMinusMax}
f(\xb)=&\max\left\{\kb_1^{\top} \xb + c_1, \dots,\kb_{|\Kc|}^{\top} \xb + c_{|\Kc|}\right\}\\
\nonumber
-&\max\left\{\lb_1^{\top} \xb + d_1,\dots,\lb_{|\Kc|}^{\top} \xb + d_{|\Kc|} \right\}
\end{align}
due to convexity of $g$ and $h$ \cite[III.C]{Hempel2015}. Now, standard implementations of explicit MPC use binary search trees to identify the ``active'' segment in~\eqref{eq:f}. In contrast, \eqref{eq:maxMinusMax} allows to evaluate $f$ by selecting the maximum from all affine segments of $g$ and $h$, respectively. For the given example, a comparison between these two methods shows an average reduction of evaluation times by a factor of $10$ while storage capacity is $16$ times reduced.

\section{Conclusions}
\label{sec:conclusions}
We presented a novel optimization-based procedure for the decomposition of a given PWA function into two convex PWA functions. In contrast to existing approaches, the novel procedure is always applicable and it provides two convex functions of identical complexity (in terms of the underlying partitions). The benefits of our scheme were illustrated with a case study on explicit MPC. Future research will focus on techniques to further reduce the complexities of the resulting partitions.

\section*{Acknowledgment}
Support by the German Research Foundation (DFG) under the grant SCHU 2940/4-1 is gratefully acknowledged.


\begin{thebibliography}{6}
\providecommand{\natexlab}[1]{#1}
\providecommand{\url}[1]{#1}
\csname url@samestyle\endcsname
\providecommand{\newblock}{\relax}
\providecommand{\bibinfo}[2]{#2}
\providecommand{\BIBentrySTDinterwordspacing}{\spaceskip=0pt\relax}
\providecommand{\BIBentryALTinterwordstretchfactor}{4}
\providecommand{\BIBentryALTinterwordspacing}{\spaceskip=\fontdimen2\font plus
\BIBentryALTinterwordstretchfactor\fontdimen3\font minus
  \fontdimen4\font\relax}
\providecommand{\BIBforeignlanguage}[2]{{%
\expandafter\ifx\csname l@#1\endcsname\relax
\typeout{** WARNING: IEEEtranN.bst: No hyphenation pattern has been}%
\typeout{** loaded for the language `#1'. Using the pattern for}%
\typeout{** the default language instead.}%
\else
\language=\csname l@#1\endcsname
\fi
#2}}
\providecommand{\BIBdecl}{\relax}
\BIBdecl

\bibitem[Bemporad et~al.(2002)Bemporad, Morari, Dua, and
  Pistikopoulos]{Bemporad2002}
A.~Bemporad, M.~Morari, V.~Dua, and E. Pistikopoulos, ``The explicit linear
  quadratic regulator for constrained systems,'' \emph{Automatica}, vol.~38,
  no.~1, pp. 3--20, 2002.

\bibitem[Nguyen et~al.(2017)Nguyen, Gulan, Olaru, and
  Rodriguez-Ayerbe]{nguyen2017convex}
N.~A. Nguyen, M.~Gulan, S.~Olaru, and P.~Rodriguez-Ayerbe, ``Convex lifting:
  Theory and control applications,'' \emph{IEEE Transactions on Automatic
  Control}, vol.~63, no.~5, pp. 1243--1258, 2017.

\bibitem[Horst and Thoai(1999)]{horst1999dc}
R.~Horst and N.~V. Thoai, ``D{C} programming: Overview,'' \emph{Journal of
  Optimization Theory and Applications}, vol. 103, no.~1, pp. 1--43, 1999.

\bibitem[Kripfganz and Schulze(1987)]{Kripfganz1987}
A.~Kripfganz and R.~Schulze, ``Piecewise affine functions as a difference of
  two convex functions,'' \emph{Optimization}, vol.~18, no.~1, pp. 23--29,
  1987.

\bibitem[Hempel et~al.(2015)Hempel, Goulart, and Lygeros]{Hempel2015}
A.~B. Hempel, P.~J. Goulart, and J.~Lygeros, ``Inverse parametric
  optimizationwith an application to hybrid system control,'' \emph{IEEE
  Transactions on Automatic Control}, vol.~60, no.~4, pp. 1064--1069, 2015.

\bibitem[De~Loera et~al.(2010)De~Loera, Rambau, and
  Santos]{de2010triangulations}
J.~A. De~Loera, J.~Rambau, and F.~Santos, \emph{Triangulations Structures for
  algorithms and applications}.\hskip 1em plus 0.5em minus 0.4em\relax
  Springer, 2010.

\end{thebibliography}
\end{document}